\documentclass[a4paper]{aa}
\usepackage{psfig}
\usepackage{epsfig}
\def \sax {BeppoSAX}
\def \src {MR\thinspace 2251-178}

\def \hcm {\hbox {\ifmmode $ atom cm$^{-2}\else atom cm$^{-2}$\fi}}
\def \arcmin {\hbox{$^\prime$}}

\def\approxgt{\mathrel{\hbox{\rlap{\lower.55ex \hbox {$\sim$}}
        \kern-.3em \raise.4ex \hbox{$>$}}}}
\def\approxlt{\mathrel{\hbox{\rlap{\lower.55ex \hbox {$\sim$}}
        \kern-.3em \raise.4ex \hbox{$<$}}}}

\begin{document}
\voffset -1.5cm


\title{BeppoSAX spectroscopy of \src: a test for 
ionized reflection in radio quiet QSOs }

\author{A. Orr\inst{1,2}
      \and P. Barr\inst{1}
        \and M. Guainazzi\inst{3}
        \and A.N. Parmar\inst{1}
        \and A.J. Young\inst{4}
}
\offprints{A. Orr (aorr@astro.estec.esa.nl)}

\institute{Astrophysics Division, Space Science Department of ESA, ESTEC,
        Postbus 299, 2200 AG Noordwijk, The Netherlands 
\and    
        Institut f\"{u}r Astronomie und Astrophysik T\"{u}bingen, 
        Waldh\"{a}user Str. 64, D-72076 T\"{u}bingen, Germany 
\and
        XMM SOC, VILSPA, Villafranca del Castillo, Spain
\and 
        Department of Astronomy, University of Maryland, 
        College Park, MD 20742, U.S.A.}

\date{Submission date: 2000 March 17; Received date: 2000 March 20 ; Accepted date: 2001 July 18}

\authorrunning{A. Orr et al.}
\titlerunning{BeppoSAX spectroscopy of \src}


\abstract{We present the 0.1--200 keV \sax\ spectrum of \src\
observed at two epochs in 1998 separated by 5 months.
Both epochs show identical spectral shape and X-ray flux. 
Analysis of the combined spectra allow us to
confirm the presence of the ionized Fe K$\alpha$ line
detected by ASCA and to test the presence of 
reflection from ionized material. 
 The best self-consistent spectral fit is obtained when including a 
contribution from a mildly ionized reflector 
($\xi_{0.01-100} \sim 1625 $ erg cm s$^{-1}$, i.e. U$_{0.1-10} \sim 0.14$) 
with a reflection fraction
R$_{\rm ion} \sim 0.11$. 
An exponential cut-off to the direct power-law continuum is 
then required at ${\rm E} \sim 100$ keV.
The low energy spectrum is absorbed by ionized matter
with a column density  N$_{\rm W} = (8.1_{-1.2}^{+2.1}) 
\times 10^{21}$ cm$^{-2}$ and an X-ray  ionization parameter 
 U$_{0.1-10}= 0.06 \pm 0.01$.
The warm absorber column is slightly lower
than that measured with ASCA. This change could be caused
by bulk motion. The BeppoSAX absorber ionization parameter globally
agrees with the ``U versus Flux'' correlation found for \src\ with 
EXOSAT, {\it Ginga} and ASCA. This suggests that the
absorber is in a state of instantaneous ionization equilibrium.
\\ } \maketitle

\section{Introduction}
\label{sect:intro}

The large variety of observed spectral and temporal variability properties 
of AGN is at the origin of the complex AGN classification scheme. 
Attempts to find a simple explanation for the different classes have
resulted in the ``unified model'' of AGN (Antonucci \cite{an:93}), in which the
viewing angle of the observer is the key parameter. 
AGN luminosity is an additional parameter which can directly influence 
the physical state of  matter located in the innermost regions
of the nucleus.
However the role of  AGN luminosity is at present still not well
understood. For instance, recent X-ray studies are still inconclusive about 
the existence of any fundamental differences  between Radio Quiet (RQ) QSOs 
and RQ Seyfert 1  galaxies (Sy~1).  These two classes have often been regarded 
as high-- and low--luminosity manifestations of a same  phenomenon 
(e.g. Staubert \& Maisack \cite{s:96}; Lawson \& Turner \cite{l:97}). 
For example, 2--10 keV power--law spectral slopes of low--$z$ RQ QSOs have been found 
to be similar to those of low luminosity Sy 1s
($\Gamma \sim 1.9-2$,
e.g. George et al. \cite{g:00}, Reeves \& Turner \cite{r:00}).
Variability and warm absorber properties appear to be comparable 
in RQ QSOs and Sy 1s (George et al. ibid.).
However,  the strong Compton reflection ``hump'' which is
common  in Sy 1s has not  been clearly detected in luminous RQ QSOs
(Williams et al. \cite{w:92}, 
Lawson \& Turner \cite{l:97}, Reeves et al. \cite{r:97}, George et al. \cite{g:00}, Reeves \& Turner \cite{r:00}).
An analysis of ASCA data for high $z$ RQ QSOs (Vignali et al. \cite{v:99})
shows that the 2--30 keV (source frame) spectra are well described
by a single power-law.  
Nandra et al. (\cite{n:97b}) showed that  Fe K$\alpha$ line emission in RQ QSOs is
 weaker  than in  Sy 1s  and that the line profile and centroid energy
 both seem to 
depend on the AGN luminosity. This trend has recently been confirmed 
by George et al. (\cite{g:00}).
Several mechanisms have been proposed to explain the differences between RQ QSOs and Sy 1s.
For instance relativistic beaming could enhance the underlying X-ray continuum emission 
but not the iron line, 
which is thought to arise by fluorescence from cold or hot reflecting 
material (Williams et al. \cite{w:92}).
Another interpretation is that  the X-ray luminosity of the central 
source
regulates the ionization state of the Compton scattering medium which
is probably the accretion disk itself. As the  luminosity increases the
atoms in the disk become increasingly ionized. At high ionization states the iron line
is emitted at higher energies and its flux may decrease (Nandra et al. 
\cite{n:97b} and references therein;  Matt \cite{m:98}; Ross et al. 1999, hereafter \cite{r:99}). 
Ultimately, 
the optical depth for photo-electric absorption decreases to such an extent 
that the continuum flux $<$30 keV is mostly reflected from the ionized 
disk and no longer absorbed (Basko et al. \cite{b:74}; 
\cite{r:99}).
Narrow band X-ray observatories have failed to detect any ``hard tails''
in RQ QSOs
and the reflection component in these sources may so far have been confused
with the underlying continuum emission. 
However, {\em at high X-ray energies the cut-off due to Compton recoil 
and Klein-Nishina effects may be detectable.}

In this paper we present broad band BeppoSAX spectroscopy of a RQ QSO.
We show that this source  has an ionized iron line 
and  exhibits a high energy cut-off, despite lacking a Compton reflection hump.
We compare our data with the predictions of reflection models
in order to test for the presence of an ionized accretion disk.

\src~ is a nearby (z = 0.064)  X-ray bright RQ QSO 
(L$_X \sim 10^{45}$ erg s$^{-1}$). 
Early X-ray observations of this source with the {\it Einstein} Observatory 
revealed both a soft excess and variable intrinsic absorption. This was 
the first detection of a  warm absorber in an AGN (Halpern \cite{h:84}).
The source  was later monitored by EXOSAT between 1983 and 1984 with a 
sequence of 15 observations (Pan et al. \cite{p:90}). 
During this period the 2--10 keV flux 
varied between 1.8 and 3.3 $\times 10^{-11}$ erg s$^{-1}$ cm$^{-2}$. 
The changes occurred on time-scales as short as $\sim$10 d. 
In the same period the column density of the warm absorber also varied. 
Strong correlations between the soft excess flux, 
the warm absorber column density variability and the source luminosity were found.
A subsequent reanalysis of the EXOSAT data together with {\it Ginga} observations
(Mineo \& Stewart \cite{m:93}) showed that the spectral behavior can be well 
described using a self-consistent warm absorber model, and that the ionization
parameter of the gas is clearly correlated with the continuum flux. 
ASCA observations in 1993 and 1996 (Otani et al. \cite{o:97};
Reeves et al. \cite{r:97}) reveal a weak emission
line at $6.57_{-0.07}^{+0.09}$ keV due to partially ionized iron. 
In this case too, the  ionization state of the warm absorber correlates well 
with the measured X-ray flux. High ionization absorption lines 
have been observed in the UV by Monier et al. (\cite{mo:01}), these appear to
be related to the X-ray warm absorber.

The BeppoSAX observations of \src\  extend 
the spectral range of previous X-ray spectroscopy of this object. 
At the same time they  also expand the luminosity range of 
``warm-absorbed'' AGN observed by BeppoSAX. A preliminary analysis of the
warm absorber data is given by Orr et al. (\cite{o:00}).


\section{Observations}
\label{sect:obs}

We present results obtained with
the Low-Energy Concentrator Spectrometer (LECS;
0.1--4~keV; Parmar et al. \cite{p:97}), the Medium-Energy Concentrator
Spectrometer (MECS; 1.8--10~keV; Boella et al. \cite{b:97}),
the High Pressure Gas Scintillation Proportional Counter
(HPGSPC; 4--120~keV; Manzo et al. \cite{m:97}) and the Phoswich
Detection System (PDS; 15--300~keV; Frontera et al. \cite{f:97}) 
on-board \sax. 

All these instruments are co-aligned and collectively referred
to as the Narrow Field Instruments, or NFI.
The MECS consists of two grazing incidence
telescopes with imaging gas scintillation proportional counters in
their focal planes. The LECS uses an identical concentrator system as
the MECS, but utilizes an ultra-thin entrance window and
a driftless configuration to extend the low-energy response to
0.1~keV. The non-imaging HPGSPC consists of a single unit with a collimator
that remained on-source during the entire observation. The non-imaging
PDS consists of four independent units arranged in pairs each having a
separate collimator. Each collimator was alternatively
rocked on- and off-source during the observation.

The region of sky containing \src\ was observed by \sax\ at two epochs:
from 1998 June 14 17:58 UT to June  18 05:33 UT and
from 1998 November 12 21:23 UT to November 16 05:32.
Good data were selected from intervals when the elevation angle
above the Earth's limb was $>$$4^{\circ}$ and when the instrument
configurations were nominal, using the SAXDAS 2.0.0 data analysis package.
The standard PDS collimator dwell time of 96~s for each on- and
off-source position was used together with a rocking angle
of 210\arcmin.
LECS and MECS data were extracted centered on the position of \src\ 
using radii of 8\arcmin\ and 4\arcmin, respectively.
The June exposures in the LECS, MECS, HPGSPC, and PDS instruments 
are 60.3~ks, 82.7~ks, 53.3~ks, and 41.1~ks, respectively.
The November exposures
are in the same order 30.5~ks, 61.2~ks, 13.9~ks, and 35.2 ~ks.


Background subtraction for the LECS and MECS were performed using the standard 
background files. For the HPGSPC it was 
carried out using data obtained when the instrument
was observing the dark Earth.
Finally the PDS background was estimated
from the offset field according to the standard  procedure.

\begin{figure*}
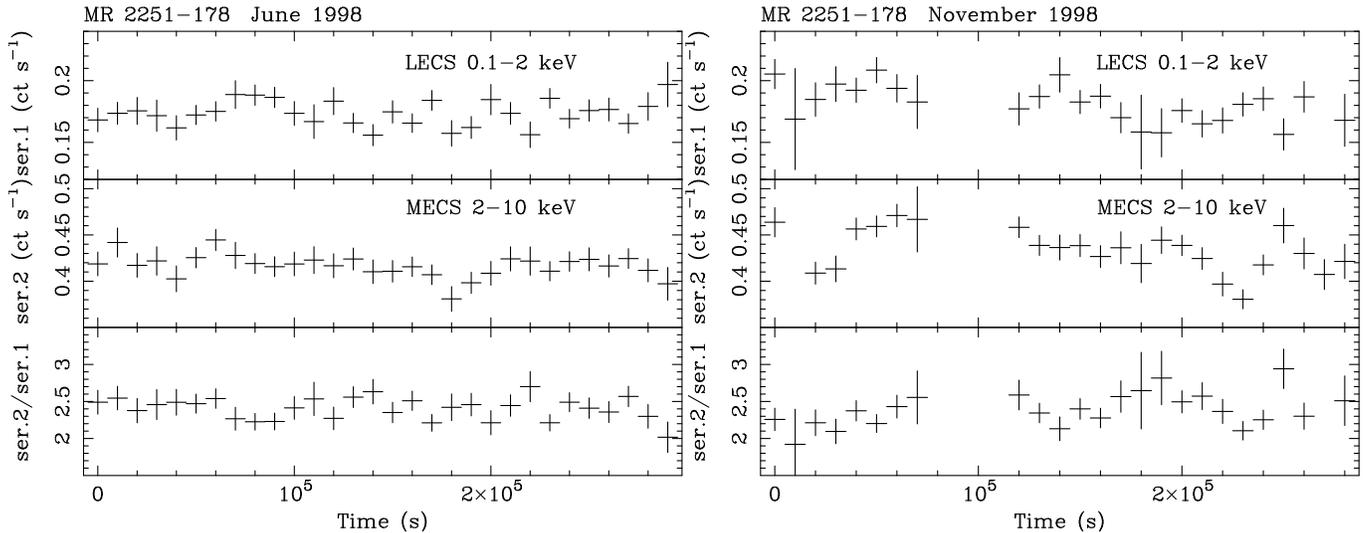

 \centerline{\psfig{figure=h2106f1.ps,width=7.0cm,angle=-90}\psfig{figure=h2106f1b.ps,width=7.0cm,angle=-90}}
  \caption[]{0.1--2 keV and 2--10 keV light curves and hardness ratios
of the June and November 1998 \sax~observations of \src~  }
  \label{fig:lightcurve}
\end{figure*}

\section{Results}

\subsection{Light curves and variability}

We have examined the LECS 0.1--2 keV and the MECS 2--10 keV light curves 
for the 1998 June and November observations as well as the corresponding 
hardness ratios. The data were binned into intervals of 10 ks.
The count rates and the hardness ratios are both relatively similar 
at the two epochs. 
The 1998 June light curves are well fit with a
constant count rate. This gives, respectively, for the soft and hard 
X-ray bands $0.173\pm 0.003$  and $0.417\pm 0.004$ counts s$^{-1}$ 
with corresponding fit statistics $\chi^2_{\nu}= 1.1$  and  0.9 (degrees of freedom, 
dof=29). 
Fits to the November data give  $0.183\pm 0.004$  and $0.433\pm 0.005$
counts s$^{-1}$, respectively, with $\chi^2_{\nu}= 1.5$ (dof=24) and 
$\chi^2_{\nu}= 3.4$ (dof=21).
The poor statistics of the constant fits to the  November data are an
indication of variability. In fact, the November light curves are 
better fitted with slowly decreasing linear functions (minus $\sim$15\%
in 70 hours). 

Finally, an interval of 30 ks beginning 10 ks after 
the observation start was ignored in the MECS November light curve. This
time interval covers a drop in MECS count rate due to a change in pointing 
which could not be corrected by using the standard housekeeping data.
The cleaned and linearized MECS November image shows what appears to
be an extended, oblong source at the position of \src. A comparision of 
light curves for different event extraction radii shows that this is
an artefact caused by the displacement of the source in the MECS field
of view. A similar displacement occurred in the LECS, however the
available housekeeping data allow this effect to be corrected.  
 
The 1998 June 2--10 keV flux is  $(4.03_{-0.12}^{+0.20}) \times 10^{-11}$ erg s$^{-1}$ cm$^{-2}$ (MECS)  and the November
flux is $(4.19 \pm 0.42) \times 10^{-11}$ erg s$^{-1}$ cm$^{-2}$.


\subsection{Spectral fits and interpretation}
\label{subsect:spectrum}

\begin{table*}
\caption[]{Fits to the summed June and November 1998  spectrum. 
The fitted energy ranges for the different instruments are
0.1--4 keV (LECS), 1.8--10 keV (MECS), 8-12 keV (HPGSPC) and 15--200
(PDS). All models include galactic
neutral absorption N$_{\rm H} = 2.78 \times 10^{20}$ cm$^{-2}$
and, apart from the simple power-law ({\tt PL}), also 3 absorption edges. 
The edge parameters are only listed for the cut-off power-law ({\tt COPL}).
All energies are given in the source rest-frame.
The {\tt pexriv} model is composed of both the reflected and
the direct component.  In this model the temperature of the ionized reflector
has been fixed at $10^6$ K.
 The parameters without uncertainty intervals were frozen during the fit.
The ionization parameter is defined as in RFY99, integrating over
the energy range 0.01-100 keV. 
The units of the ionization parameter $\xi$ are erg s$^{-1}$ cm}
\begin{center}
\begin{tabular}{lllllll}
\hline\noalign{\smallskip}
\noalign{\smallskip\hrule\smallskip}
model & $\Gamma$ &  E$_{\rm Fe K\alpha}$, E$_{\rm cut}$  & EW & E$_{\rm Edge}$ & $\tau$ & $\chi^2_{\nu}$ (dof) \\ 
      & &  keV & eV &  keV & & \\  \hline\noalign{\smallskip}
{\tt PL}   &  $1.61\pm 0.01$ &  & & & &  4.23 (162) \\
{\tt PL}$+$FeK$\alpha$  &  $ 1.67\pm 0.01$ & $6.54_{-0.10}^{+0.11}$ & $ 89\pm 24$ &  & & 1.22 (156)\\
{\tt COPL}$+$FeK$\alpha$ &   $ 1.63\pm 0.02$ & $6.54
_{-0.11}^{+0.12}$ & $ 79\pm 24 $ & 0.74 & $ 0.59_{-0.14}^{+0.15}$ &  0.94 (155)\\
                &    &  $ 133_{-35}^{+64}$  &     & 0.87 & $ 0.33 _{-0.13}^{+0.12}$ & \\
                &    &    &     & $ 1.32 _{-0.07}^{+0.08}  $ & $ 0.20_{-0.05}^{+0.06}$ & \\

\noalign{\smallskip\hrule\smallskip}

 model  & $\Gamma$ & E$_{\rm Fe K\alpha}$, E$_{\rm cut}$  & EW & R  & $\xi$ & $\chi^2_{\nu}$ (dof) \\ 
    \noalign{\smallskip\hrule\smallskip}
{\tt pexrav}$+$FeK$\alpha$  & $1.63\pm 0.02 $ &
   $6.54_{-0.11}^{+0.13}$ & $73_{-23}^{+24} $ & $<0.38 $  & &   0.93 (154) \\

{\tt pexriv}$+$FeK$\alpha$  & $1.62_{-0.02}^{+0.01}$ & $ 6.53_{-0.12}^{+0.14}$ &  $62_{-25}^{+12}$ &  $ 0.16_{-0.09}^{+0.15}$ &  $465.1_{-240}^{+2003}$ & 0.89 (153) \\
                                &                  & $115_{-30}^{+38}$  &   &  \\ 

{\tt \cite{r:99}} & $1.58\pm 0.03$ & $102_{-26}^{+39}$ &  & $0.11_{-0.05}^{+0.06}$ & $1625 _{-930}^{+1422}$ & 0.94 (155) \\
 \noalign{\smallskip\hrule\smallskip}
\end{tabular}
\end{center}
\label{tab:aver_fits}
\end{table*}

\begin{figure}
  \centerline{
  \hbox{
   \psfig{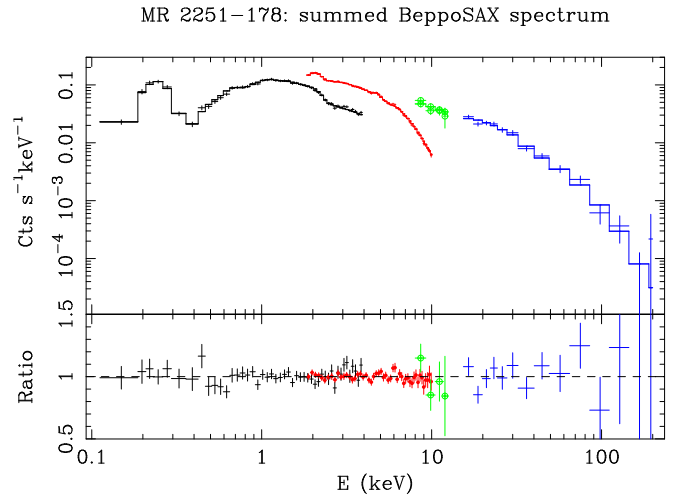}
}
}
  \caption[]{The summed June and November 1998 BeppoSAX spectra of \src. 
The MECS  data are represented by small filled squares, 
the HPGSPC data by open circles. The fit model is RFY99 with a 
warm absorber}
  \label{fig:spectra}
\end{figure}

\begin{figure}
  \centerline{
  \hbox{
   \psfig{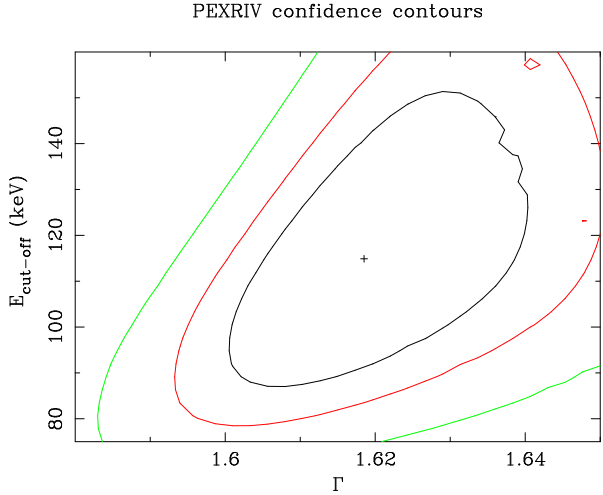}
}
}
  \caption[]{Contour plot of the parameters E$_{\rm cut-off}$  and
  $\Gamma$ in the ionized reflection model {\tt pexriv}. The contours
  indicate 68, 90 and 99\% confidence levels
}
  \label{fig:contours}
\end{figure}

\begin{figure}
  \centerline{
  \hbox{
   \psfig{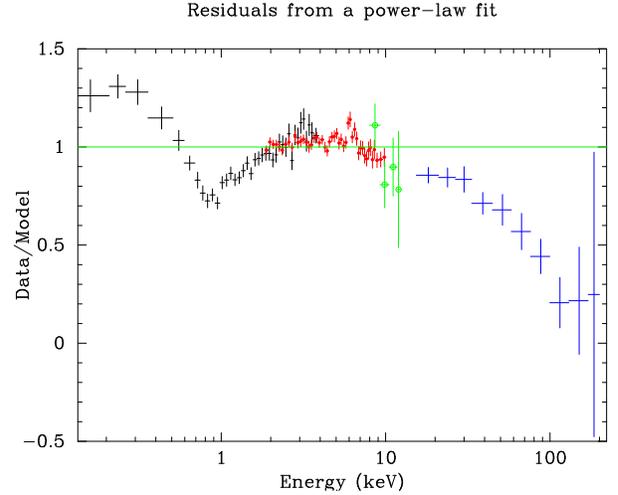}
}
}
  \caption[]{Data-to-model ratio for a simple power-law with galactic neutral 
   absorption. The MECS data are
   shown with solid squares and the HPGSPC with open circles. The 
   warm absorption and the high energy cut-off are evident
}
  \label{fig:residuals}
\end{figure}

The June and November spectra of \src\ were  investigated both individually and as a 
single, summed spectrum by simultaneously
fitting data from all the \sax\ NFI with the help of the {\tt XSPEC} 11.01
spectral fitting package.
The LECS and MECS spectra were rebinned to oversample the full
width half maximum (FWHM) of the energy resolution by
a factor 3 and to have, additionally, a minimum of 20 counts 
per bin to allow use of the $\chi^2$ statistic. 
The HPGSPC spectrum  was rebinned to have a minimum of 40 counts per bin 
and the PDS data were rebinned using the standard logarithmic binning recommended for this instrument.
Data were selected in the energy ranges
0.1--4.0~keV (LECS), 1.8--10~keV (MECS), 8.0--12~keV (HPGSPC),
and 15--200~keV (PDS) 
where the instrument responses are well determined and sufficient
counts obtained. 
The photo-electric absorption
cross sections of Morrisson \& McCammon (\cite{m:83}) and the
solar abundances of Anders \& Grevesse (\cite{a:89}) are used throughout.

Factors were included in the spectral fitting to allow for flux normalization 
uncertainties between the instruments. For all instruments factors were constrained to be within their usual ranges during the fitting (see Fiore et al. \cite{f:99}).   
In this paper all uncertainties are 90\% confidence intervals for one
parameter of interest, unless stated otherwise.


Because of the lower exposure times during the November 1998 observation the 
uncertainties on the fitted parameters are slightly larger than for those of the June spectrum.
However, our fit results indicate that {\em absolutely  no} significant spectral
change occurred between the two epochs which are separated by $\sim$5 months.
{\em All} the fit parameters for the individual spectra are compatible within the 
statistical uncertainties for {\em any} of
the models tested. For this reason and because the observed fluxes in June and November are
nearly identical with no large structure change in the light curves we have summed the spectra
for the two epochs in order to obtain a higher signal to noise ratio.


We tried fitting the summed continuum with several models (see Table 
\ref{tab:aver_fits}), starting with a power-law 
which gives a totally unacceptable fit 
($\chi^2_{\nu} = 4.2$, dof=162). Strong deviations appear in the ranges
E$\sim$0.4--3 keV and at E$\approxgt$ 80 keV as illustrated in Fig. 
\ref{fig:residuals}. The spectral complexity in the
soft X-rays is due to the warm absorber and is discussed further on. 
At high energies the data points clearly fall below a simple power-law.
Highly significant improvement is brought with an exponential 
high energy cut-off to the power-law at E$_{\rm cut-off}> 70$ keV, 
(the confidence level of the improvement is $>$99.9\%; 
F-statistic F$>$40 for $\Delta $dof=1, with respect to a 
simple power-law, with or without an Fe K$\alpha$ line or a warm absorber;
the value of the cut-off energy is discussed below). 
However, as described in the following sections,
the best fit to the summed 0.1-200 keV 
BeppoSAX spectrum of \src\ is obtained with an ionized reflection model. 
Such a model predicts a drop in high X-ray flux but, unlike the case of
reflection onto cold matter, a weaker or absent ``Compton hump''. 

\subsubsection{Significance of the high-energy cut-off}

The knowledge of the spectral shape above $\sim$10 keV is an important key 
to the interpretation of the present \sax\ data. Therefore we have taken 
special 
care in verifying the quality of the PDS data (for details about the PDS 
instrument see Frontera et al., \cite{f:97}).

It has been shown (by Guainazzi and Matteuzzi, \cite{gu:97})
that the total systematics in the PDS 20--200 keV range
due to background subtraction are lower than 0.026 s$^{-1}$
(corresonding to approximately 5.8 \% of the PDS 20--200 keV 
source count rate for \src, which is 0.45 cts s$^{-1}$). The observed flux of
a one mCrab source with photon index 1.5--2, 
is well above the systematics ($\gg 1 \sigma$) up to 60 keV and remains 
above the systematics ($\geq 1 \sigma$) up to 100 keV).
(\src\ has a flux of $\sim$3 mCrab in the 20--200 keV range).

We have also checked the consistency of the spectra measured by the
two PDS half-arrays 
(A \& B) which see the sky (source and background field) through two 
different collimators. This was done by fitting either one of the PDS
half-arrays (A \& B) spectra together with the LECS, MECS and HPGSPC 
data. Because the data were obtained during two epochs (June and 
November 1998), we first tested the half array data for June only 
and then for the summed
June and November. This is justified because the PDS rocking 
collimator positions happen to coincide within 0.6 degrees. 
Using the ``cut-off power-law + 3 edges + Fe K$\alpha$ line'' model 
in Table \ref{tab:aver_fits} one finds that the cut-off energies 
are all compatible with those
found using the combined PDS data (A + B half-arrays). Also, in all 
cases the presence of the cut-off is if significant at greater than the
99.5\% confidence level using the F-statistic. 

We have also tested whether the PDS so-called``spike'' events can
contribute to the PDS spectral shape. The answer is no: in our
two observations of \src~the contribution of the  ``spike'' events 
to the PDS spectral shape is negligible.

However, PDS event filtering using the so-called ``Variable Rise Time'' 
(VRT) selection does bring a significant improvement to the quality of 
fit, by increasing the signal to noise ratio of the PDS data. Therefore 
this filtering has been applied to all our PDS data.

\subsubsection{The {\tt pexriv} model}

Our results with a {\tt pexriv} model including a warm absorber and an Fe
K$\alpha$ line (see also Fig. \ref{fig:contours}) are the following:
a power-law slope $\Gamma= 1.62 _{-0.02}^{+0.01}$, 
a high energy exponential cut-off of the
incident power-law at E$_{\rm cut-off}= 115_{-30}^{+38}$ keV, 
the temperature of the 
ionized reflector T$_{\rm disk}$ being fixed at $10^6$ K, 
an ionization parameter of the reflector 
$\xi=465_{-240}^{+2003}$ erg cm s$^{-1}$ and
finally the scaling factor for reflection R$_{\rm ion}=0.16_{-0.09}^{+0.15}$,
which is in fact the solid angle 
R=$\Omega/2\pi$ subtended by the reflecting disk.
It should be noted that, for these reflection models, we use $\xi$ as it is 
defined in \cite{r:99}, i.e. with the ionizing flux integrated between 0.01--100 keV.
Because in the {\tt pexriv} model $\xi$ is defined
between 5--20 keV we have made the necessary
conversions, assuming a photon index $\Gamma = 1.6$
(i.e., $\xi_{\rm RFY99} \sim \xi_{\rm pexriv}\times 4.4$).

The fit statistics are good, with $\chi^2_{\nu}= 0.89$ for 153
dof.
The low value of the scaling factor R tells us that most of the flux
in this model is still coming from the direct, non-reflected component.
Furthermore, one should bear in mind that a moderate ionization 
parameter such as obtained with this model, still produces a significant 
reflection ``hump''. 
Since the present data obviously lack such a hump the scaling factor 
cannot be large.
In all our reflection fits we fixed the inclination angle of the disk with
respect to the observer so that  $\cos i = 0.45$. This was necessary because
in our fits the viewing angle and the reflection scaling factor are not 
independent, nor are they well constrained with respect to one another
at low values (R$\approxlt 0.3$ or $\cos i \approxlt 0.2$).
In fact, in the present {\tt pexriv} fit the entire  range of numerically 
permitted $\cos i$ values, i.e.  $\cos i = 0.05- 0.95$ lies within the 
99\% confidence interval for this parameter. At $\cos i \sim 0.05$ R is poorly
constrained with a lower limit value of R$>0.4$. At $\cos i \sim 0.95$ we have
$0.05<{\rm R}< 0.22$.
R and $\cos i$ are linked in the fitting process because smaller values 
of $i$ produce stronger reflection humps
and vice-versa.

\subsubsection{The {\tt pexrav} model}

Reflection on cold matter (e.g.  neutral Compton reflection model
{\tt pexrav} in {\tt XSPEC}) gives, compared to {\tt pexriv}, a slightly worse
fit  with $\chi^2_{\nu}= 0.93$ for 154. The difference is significant at 
the 99\% confidence level. 
The neutral reflection scaling factor is weakly constrained with an upper limit 
R$_{\rm neut.}<0.38$. 

\subsubsection{A self-consistent ionized reflection model: {\bf RFY99}}

The {\tt pexriv} model in {\tt XSPEC} does not include Fe K$\alpha$
emission and one therefore needs to add a separate Gaussian line component
to the fit model. 
Because of this drawback we have also tested the self-consistent 
ionized reflection models published by Ross and collaborators (\cite{r:99})
which include Fe K$\alpha$ line emission and take into account the destruction
of  K$\alpha$ photons by Auger effects.  
The models were specially computed for the values of the incident 
power-law slope observed in \src.
As with  {\tt pexriv}  we set the iron abundance at the solar 
value and included a warm absorber. The RFY models do not include the
``incident'' continuum  so it is necessary to add
it separately, in this case  an exponentially 
cut-off power-law. We hereafter refer to this combination of components
as the ``RFY99 model''.
The power-law slope of the incident continuum and the slope specifying the 
reflection component were fixed to have the same value.  
It should be noted that, unlike {\tt pexriv}, the RFY99 reflection
assumes a sharp, step-like cut-off 
\footnote{As mentionned in the RFY99 paper, 
extending the illuminating spectrum to higher energies would 
raise the corresponding portion of the emergent spectrum
via Compton downscattering of higher energy photons. }
of the incident spectrum at exactly 100 keV. This causes the RFY99 reflection
component to decline steeply above $\sim$50 keV.
The RFY99 model gives a good a fit with
$\chi^2_{\nu} =  0.94$ for 155 dof , see Fig.\ref{fig:spectra}. 
If the  incident spectrum is taken to be a power-law without an exponential
cut-off the fit becomes significantly worse ($\chi^2_{\nu} =  1.40$).
The ``RFY99'' reflected fraction is $\sim$0.11 and is compatible with the {\tt pexriv} value. 
Both models also give comparable high energy cut-offs
at $\sim$100 keV. We have checked  that this value is {\em not} an artefact caused by the sharp cut-off
assumed in the RFY99 reflection component (at E$_{\rm rest} = 100$ keV and 
E$_{\rm obs} = 94$ keV). Indeed, 
if a ``hard tail'' is added to the  RFY99 reflection component, with an exponential
cut-off at 200 keV, the fitted continuum cut-off energy remains at 
$\sim$100 keV with a 90\%
confidence upper limit E$_{\rm cut}<$200 keV. The reason for this is most likely the relatively low reflection scaling factor, R$\sim$0.11.

A noteworthy analogy can be seen between the {\tt pexriv} and \cite{r:99} models concerning the
ionization parameter of the reflector. Due to non-linear $\chi^2$ space
two distinct fits are possible with the RFY99 models. In the
first case the ionization parameter is $\xi=1625_{-930}^{+1422}$ erg cm s$^{-1}$ 
and in the second $\xi < 35.4$  erg cm s$^{-1}$. This second value corresponds
to a nearly neutral reflector and gives a worse fit than with the first value
($\Delta \chi^2 = 4.4$). Such a behavior is  
analogous to the {\tt pexriv} versus {\tt pexrav} solutions and therefore
gives support to the validity of the ionized reflection models and their fitted parameter values.

These reasons lead us to consider hereafter only the fit solution yielding the
higher value of $\xi$. The ionization parameter measured with  {\tt pexriv} is
poorly constrained making it difficult to compare with the 
RFY99 value.  

The value of $\xi$ (with the RFY99 model + 3 edges) 
is  driven mainly by the two following factors:
first, the intensity of the Fe K$\alpha$ line 
 and second,
the low energy ($<$6.4 keV) spectral shape.
This is verified by performing the fits over limited energy 
ranges. For instance, excluding the 5.5--8 or 5--12 keV range leads to 
significantly higher ionization parameters, and weaker Fe lines. 
Likewise, if the range below 5 keV is excluded, then $\xi$
decreases significantly.

\begin{figure}
  \centerline{
  \hbox{
   \psfig{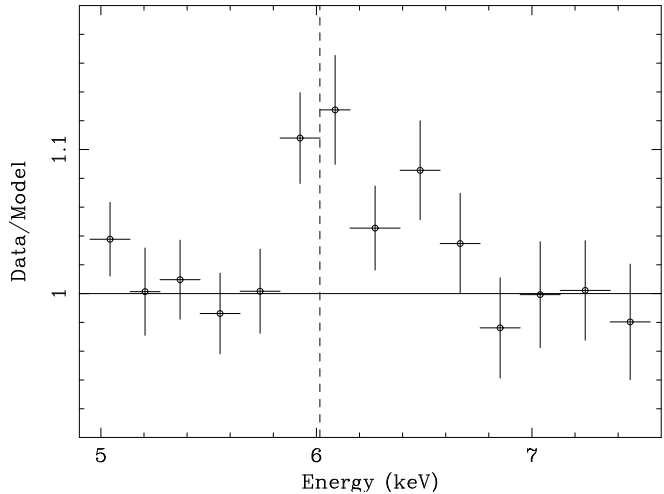}
}
}
  \caption[]{Fe K$\alpha$ line profile: 
MECS data to model ratio in observer's reference frame. The model
  is a cut-off power-law  with a Gaussian
emission line for Fe K$\alpha$, where {\em the line normalization has been set
to zero for the plot}. The energy of neutral Fe K$\alpha$ emission (6.4 keV)
is indicated by the dashed line}
  \label{fig:fe_line}
\end{figure}

$\xi=1625$ implies a hot Fe K$\alpha$ line in the reflection component,
with E$_{\rm rest} \sim $ 6.8 keV and  E$_{\rm obs} \sim $ 6.4 keV.
This is ``hotter'' than the line energies found with the other, 
non-self-consistent, fit models in Table 1. 
However, from Fig. \ref{fig:fe_line} we see that the ``hot'' line is
still consistent with the MECS data. 
Indeed, no large fit residuals appear at the Fe K$\alpha$ line
with the RFY99 model. 

Nevertheless, an improvement in $\chi^2$ can be brought to 
the {\em global} RFY99 fit if an additional Fe K$\alpha$ line
(E$_{\rm rest}= 6.53 _{-0.14}^{+0.15}$ keV, EW = $55 _{-11}^{+32}$ eV)
is included ($\chi^2_{\nu}=0.9$,  $\Delta \chi^2 = 8.8$ 
for $\Delta$(dof)=2). But the 
model is then no longer self-consistent and, in order to ``compensate'' 
for the additional line, the ionization parameter of the reflector tends to 
increase and the reflected fraction to decrease.
It may be a hint that a slightly different Fe abundance is required.
For instance, increasing the Fe abundance increases both the line flux and 
the depth of the Fe K-edge. However, any iron over-abundance must be 
relatively low.
This is because the ionization parameter derived from the RFY99 model
indicates that the reflecting matter is in the regime where
ions are already too highly ionized to permit Auger destruction of iron
line photons. Therefore these photons escape from the reflector to
produce the observed hot iron line (Fabian et al. \cite{fa:00}).
Another possibility is that the line is due to neutral reflection from matter
further out, e.g. the torus or a cold phase of the accretion disk.

%

\subsubsection{The Fe K$\alpha$ line}

Fits of the Fe K$\alpha$ line (see Fig. \ref{fig:fe_line}) 
show that it is weak but very significant at $>$99.9\% confidence level
(the F-test value for the presence of a line in the cut-off 
power-law model of Table \ref{tab:aver_fits} is ${\rm F}=15.9$ for 
a difference of 2 dof).
A narrow line fit gives E$_{\rm line}({\rm rest}) = 6.53_{-0.12}^{+0.14}$ 
keV, an equivalent width ${\rm EW} = 62 _{-25}^{+12}$ eV 
(the continuum is here the best fit {\tt pexriv} model). 
The line energy is consistent with ionization stages Fe {\sc xiii-xxiii}. 
A significantly worse fit is obtained if one forces the line energy to be
6.4 keV ($>$95 \% confidence level).
In the fit models of Table \ref{tab:aver_fits}
the width of the narrow line has been frozen at 
$\sigma = 0.1$ keV because no there is no significant 
improvement in the fit statistics when leaving $\sigma$ free.

Nevertheless, we checked whether broad wings are present beneath 
the narrow core.  
We fitted the MECS continuum spectrum with a power-law continuum and 
galactic absorption to which we added, in turn, a narrow line 
or a ``narrow plus broad Gaussian line''. In the second case 
both lines were required to have the same centroid energy.
There is no strong evidence for the presence of a broad Gaussian component of 
the Fe K line. Such a broad line is
not required by the data, the confidence level for its inclusion
being only $\sim$80\%, using the F-statistic. The resulting line 
width is $\sigma_{\rm broad}$ = 1.25$^{+2.22}_{-0.77}$ keV.
The equivalent widths of the core and the broad components are then
53.0 and 116.0 eV, respectively. So, we cannot completely 
rule out some broadening of the line by e.g. Compton scattering.

Likewise, we have performed fits with the {\tt XSPEC diskline} model using
the MECS data alone and compared them with the broad gaussian line fits. 
The fit quality of either model is statistically
equivalent.  
Therefore it is not possible to formally exclude 
the presence of a relativistically  ``blurred'' line. 
The width of the peak line emission remains lower than the instrument's 
energy resolution at 6 keV ($\sim$8\% in the MECS).

When testing the {\tt diskline} model we tried to minimize the amount of
free parameters by estimating the outer radius of the disk.
We considered the simple case of an annulus illuminated from one 
side and a non-rotating central black hole 
with a mass of $2.5 \times 10^8\;{\rm M}_{\odot}$. This mass was derived for \src~by 
Brunner et al. (\cite{br:97}) by fitting geometrically
thin $\alpha$ accretion disc models. They also found an inclination of the 
disc of 70 (39--82) degrees.
From the value of the ionizing X-ray flux and the ionisation 
parameter obtained with the RFY99 fit it is possible to estimate the 
surface area of the ionized disk, assuming a hydrogen number density of
n$_0\sim 10^{15}$ cm$^{-3}$. From the value of this surface area one
can derive the outer radius of the disc ($\sim$6.46 R$_G$), the inner 
radius being fixed at 6 R$_G$. 
The resulting {\tt diskline} feature is double peaked, several 
keV broad and skewed. The blue peak is, however, 
about 4 times stronger in intensity than the red 
peak. And the FWHM of the blue peak is $<$0.1 keV (not resolved by 
BeppoSAX).

In conclusion, our results with a narrow emission line
confirm the previous measurements of the line made at higher energy 
resolution with  ASCA (E$_{\rm line} = 6.57_{-0.07}^{+0.09}$ keV,  
EW = $62 \pm 32$ eV,  $\sigma < 0.22$ keV,  Reeves et al. 1997)
and they also are consistent with the EW-luminosity relation suggested by
  Iwasawa \& Taniguchi (\cite{i:93}) and confirmed by Nandra  et al. 
(\cite{n:97b}.)
This relation is a strong anticorrelation between the luminosity and the 
EW, an ``X-ray Baldwin effect'' which suggests Fe K$\alpha$
line core EWs lower than $\sim$80 eV at 
L$_{2-10\;{\rm keV}}\approxgt 10^{45}$ erg s$^{-1}$,
which is the luminosity of \src.

\subsubsection{The complex low-energy X-ray spectrum}

A soft excess component is not required by the data.
The ionized gas of the warm absorber 
can account for the spectral complexity below 2 keV. 
Likewise, we see that the best fitting models (which include a warm absorber)
do not require  excess cold absorption. The upper limit on excess neutral 
absorption is $0.16\times 10^{20}$ cm$^{-2}$, using the RFY99 model. 

The soft X-ray spectrum of \src\  is complex and can be well fit by considering the
effects of strong absorption
by partly ionized matter. Below 1 keV this absorption is essentially due to
highly ionized oxygen. 
BeppoSAX cannot spectrally resolve the two oxygen edges O {\sc vii} and  O {\sc viii}  
at 0.74 and 0.87 keV in the source rest frame. 
Therefore fits have been performed by multiplying the continuum model by 
either a single ``blended'' absorption edge or two oxygen edges with 
energies fixed at the values above. 
The ``blended'' edge parameters are: E$_{\rm edge}= 0.81\pm 0.01$ keV,
$\tau = 0.80\pm 0.05$. In the second case, $\tau_{\rm O7}= 0.38
\pm 0.07$ and  $\tau_{\rm O8}= 0.48\pm 0.06$.
A third edge brings further significant improvement to the fit statistics
(F-statistic F=21.41 for $\Delta $dof=2).
This third edge (${\rm E} \sim 1.32$ keV, $\tau \sim 0.20 $) can be 
due to Ne {\sc x} K,  Mg {\sc i-iv} K or  Ni {\sc xiv-xv} L.  


Alternatively, fits with  warm absorber models were tested. The models
were calculated using ION98, the 1998 version of the photo-ionization code ION
(Netzer \cite{ne:96}). They were combined with a cut-off power-law continuum
and a gaussian Fe K$\alpha$ line.
The best fit gives a column density of the warm absorber N$_{\rm W} = (8.1_{-1.2}^{+2.1}) 
\times 10^{21}$ cm$^{-2}$ and an oxygen ionization parameter U$_{\rm oxy} = 0.018 _{-0.003}^{+0.004}$. 
The oxygen ionization parameter  U$_{\rm oxy}$ is the unitless
ratio of ionizing photon flux to the
electron density and is defined by integrating  the rate of photon emission over the oxygen 
K-shell continuum, i.e.  between 0.538--10~keV. In the present case U$_{\rm oxy}$ 
translates to an X-ray ionization parameter (defined by integrating 
between 0.1--10 keV) U$_{0.1-10}= 0.060_{-0.009}^{+0.012}$.
The fit statistics with the warm absorber model are significantly better (at more than
90\% confidence) than with the triple absorption edge fits: $\Delta \chi^2 = 5.13$
for $\Delta $dof=2 , leading to an F-statistic F = 2.9.

We note that the ionization parameters of the reflection component
and the warm absorber are similar: using the apropriate conversion 
and flux integration between $\xi_{0.01-100}$ and 
U$_{0.1-10}$ (e.g. Netzer \cite{ne:96},
 George et al. \cite{g:98}) the reflector has U$_{0.1-10} = 0.04 
(0.02-0.22)$, with the {\tt pexriv} fit and  U$_{0.1-10} = 0.14 
(0.06-0.27)$, with the RFY99 fit, respectively.
But the warm absorber is optically thin and the RFY99 reflector 
optically thick.  So, despite their similar ionization parameters
they most likely are distinct. 

The comparison of our warm absorber fit results with previous warm absorber
results for \src\ is complicated by the different formalisms  used in the 
literature 
to define the ionization parameter, in particular the integration range 
of the ionizing photon flux.
For a start, we note that our measured warm absorber column, N$_{\rm W} = 
(8.1_{-1.2}^{+2.1}) \times 10^{21}$ cm$^{-2}$ 
is consistent with values derived from {\it Ginga} and EXOSAT data 
(Mineo et al. \cite{m:93}), even when \src\ was in weaker or brighter 
states than seen by BeppoSAX. An earlier study of the same EXOSAT data 
by Pan et al. (\cite{p:90}) seemed to suggest a column versus flux trend
whereby the warm absorber column decreases with increasing X-ray flux. 
However, results from Mineo et al. (\cite{m:93}) as well as from Otani et al.
(\cite{o:97}), for ASCA data, excluded such a trend. Unlike the EXOSAT 
data with relatively low statistics, the high signal to noise 
ASCA spectra indicate slightly  lower values of  N$_{\rm W}$ than 
the \sax\ spectra do ($\ll7.5 \times 10^{21}$ 
 cm$^{-2}$).
Finally, the {\it Ginga}, EXOSAT and ASCA data all show a well defined correlation 
between the ionization parameter and the X-ray flux. Using the 
ionization parameter ``translation'' curves of George et al. 
(\cite{g:98}, Fig. 1) which give the relations between various 
definitions of the ionization parameter, $\xi$, U and U$_{0.1-10}$, 
we estimate that the ionization parameter of the warm absorber in
\src~globally agrees with the EXOSAT, {\it Ginga} and ASCA ``U versus flux''
correlation curves. 
To summarize, it seems that the warm absorber in \src\ is responding
in a consistent way to the slow variations of the ionizing continuum 
flux and therefore is at most times in a state of ionization \
equilibrium. Any change in absorber column density between the 1996
ASCA  and the 1999 \sax\ observations could be explained by bulk
motion of the absorber material across our line of sight.

\section{Discussion}
\label{sect:discussion}

Both the {\tt XSPEC pexriv} and the RFY99 models give very good fits to the \sax\
spectrum of \src.
Of all the models we tested only the RFY99 model can by itself and in a 
self-consistent  way  account for  the ionized Fe line, part of the
high energy cut-off and the lack of reflection hump, 
however, this model assumes a reflector with 
constant density. The upper layers of the reflector, which are subject 
to strong external illumination probably have lower densities and higher 
effective ionization parameters than the inner layers. 
This would tend to suppress Fe K$\alpha$ emission in the outer layers
(\cite{r:99}).
Nayakshin et al. (\cite{na:00}) and Ballantyne et al. (\cite{ba:01})
have recently calculated models of reflection on an ionized atmosphere in
hydrostatic equilibrium. Nayakshin et al. (\cite{na:00})
find that in certain conditions
Fe K$\alpha$ photons can nevertheless be emitted: 
{\em if} the incident spectrum is steep ($\Gamma\approxgt 2$)  
or {\it if } the 
intrinsic disk flux exceeds the X-ray illuminating flux, in these cases
the reflecting layer is expected to yield ionized iron emission lines.
Ballantyne et al.  (\cite{ba:01}) show that the constant density models 
of Ross \& Fabian (\cite{rf:93}) actually often give 
accurate fits to their ``hydrostatic'' reflection spectra, 
both statistically and parametrically, assuming an ASCA-like spectral 
resolution. On the other hand, the {\tt XSPEC pexriv} and {\tt pexrav}
models give only poor fits to their spectra.

Whereas the value of the Fe K$\alpha$  line energy, as measured by 
ASCA and \sax\ in \src, clearly excludes neutral Fe, the case for 
ionized reflection is less firm. The data strongly suggest
that little reflection is
present and that we are in fact mainly observing  the direct ionizing
continuum since very good fits are obtained with an exponentially 
cut-off power-law
($\chi^2_{\nu} = 0.94$, dof=155) or a {\tt pexriv} or \cite{r:99} model 
allowing the direct component to ``shine'' through at more than 80\% 
($\chi^2_{\nu} = 0.89$, dof=153 with {\tt pexriv}, see also Table \ref{tab:aver_fits}). Because of the weak contribution of the reflection
component the shape of the cut-off at 
$\sim$100 keV must be intrinsic to the primary X-ray emission.
However, despite its weak contribution, an ionized reflector can 
produce the observed ionized Fe emission, as shown with the RFY99 fit. 

Broad band observations of Sy1 galaxies with \sax\
have revealed exponential cut-off energies $\approxlt$100 keV
in only a couple Sy1s: 
e.g. NGC~5548 (E$_{\rm cut-off}= 115_{-27}^{+39}$ keV, Nicastro 
et al. \cite{n:00}) and NGC~4151 (70 keV, Piro et al. \cite{p:98}).
Whereas in other Sy1s the cut-off occurs at much higher values, 
with lower limits around 200--300 keV (c.f. Gondek et al. \cite{go:96}; 
Matt \cite{m:98}).
More recently,  Zdziarski et al. (\cite{z:00}) have measured the average OSSE spectra 
of Sy~1 and Sy~2. Using a cut-off power-law they derived expontenial cut-off energies 
of 120$_{-60}^{+220}$ and 130$_{-50}^{+220}$ keV 
for Sy~1 and Sy~2s, respectively. 
It appears therefore that the cut-off energy measured by \sax\ for the RQ QSO 
\src\ is compatible with the average OSSE Sy~1 value as well as  the 
\sax\ values obtained so far for Sy~1s.

\section{Conclusion}
\label{sect:conclusion}

We have for the first time observed \src\ in the 0.1--200 keV 
energy range using BeppoSAX.
The spectral fitting described above shows that this source possibly
combines two distinct signatures of ionized reflection which, to our knowledge,
have not been detected together before in an individual radio quiet AGN: 
a strongly ionized
Fe K$\alpha$ emission line and a contribution, in this case relatively weak, 
from  an ionized reflection continuum.
A spectral break is apparent at higher energies which can be parametrized with the 
help of  an e-folded power-law. The cut-off energy is measured to be in the range 
$\sim$80-190 keV (90\% confidence interval) and is compatible with values found for
Sy~1 galaxies.
\begin{acknowledgements}
The \sax\ satellite is a joint Italian-Dutch programme. 
AO acknowledges a Fellowship of the Swiss National Science Foundation
during part of this work. She expresses gratitude to Prof. R. Staubert for having hosted 
her at the IAA in T\"ubingen during this period and wishes to thank J. Wilms and
other colleagues form the IAAT for fruitful discussions about \src.
 AO is very thankful to Prof. Netzer for allowing to
use his ION98 model and giving valuable advice about the warm absorber 
fitting.
\end{acknowledgements}

\end{document}